# TC-SKNet with GridMask for Low-complexity Classification of Acoustic scene

*Luyuan Xie, Lin Yang, Zhaoyu Yan, Yinping Zhang,  Junjie Wang*

School of Software and Microelectronics, Peking University, Beijing, China

## Abstract

Convolutional neural networks (CNNs) have good performance in low-complexity classification tasks such as acoustic scene classifications (ASCs). However, there are few studies on the relationship between the length of target speech and the size of the convolution kernels. In this paper, we combine Selective Kernel Network with Temporal-Convolution (TC-SKNet) to adjust the receptive field of convolution kernels to solve the problem of variable length of target voice while keeping low complexity. GridMask is a data augmentation strategy by masking part of the raw data or feature area. It can enhance the generalization of the model as the role of dropout. In our experiments, the performance gain brought by GridMask is stronger than spectrum augmentation in ASCs. Finally, we adopt AutoML to search best structure of TC-SKNet and hyperparameters of GridMask for improving the classification performance.  As a result, a peak accuracy of 59.87% TC-SKNet is equivalent to that of SOTA, but the parameters only use  20.9 K.

## 1. Introduction

The objective of acoustic scene classifications (ASCs) is to predict specific locations based on sound events and auditory information. Some recent studies have shown that it is very effective to improve the performance of ASCs through deep learning technologies. CNNs become very popular in speech recognition [1-4] and keyword spotting [5-7]. Various CNNs' structures greatly contributed to the performance improvement of ASCs [8-10].

Hu et al. used low-complexity models (MobNet and small-FCNN) for fixed-length acoustic scenes' task with augmentation methods (mixup and spectrum augmentation). The MobNet is the improved version of  MobileNetV2 [11] , which has the low-complexity and high accuracy. Small-FCNN [10] uses fully convolutional layers and channel attention. Seo et al. adopted Conformer for low-complexity classification of variable-length acoustic scene data and achieved good performance. But based on less training data, self-attention is easy to fall into overfitting[9]. In the ASCs under low-complexity conditions, we focus on three novel technologies — reducing model complexity and computing cost by temporal convolution, improving the robustness by GridMask [12] data augmentation method, and selecting the high-quality model by AutoML.

Now there are few studies on the relationship between the length of speech and the size of the convolution kernels. In a convolutional layer, no matter what the length of the target speech is, it is mandatory to use  fixed size kernels at the same time, which is similar as in the computer vision field. Most models (VGG [13], ResNet [14], MobileNet etc.) are aimed at targets of different sizes and also use convolution kernels of the same size, but Selective Kernel Networks [15] can be targeted at the target size selects the length of the convolution kernels. This feature helps to improve performance while keeping low complexity. In order to further reduce the complexity of the model. we adopt Temporal-Convolution which not only guarantees the accuracy but also reduces the computing complexity compared to 2D-CNN. Based on these, we combined Temporal-Convolution [16] and Selective Kernel Networks  to build TC-SKNet to solve the problem of low-complexity classification of ASCs.

With the development of deep learning, more and more attention has been paid to the data augmentation technologies Data augmentation schemes are mainly divided into two types in the field of speech signal processing or speech recognition: one is focused on the manipulation on the time domain, such as adding noises, reverberations and speed perturbations on the raw data, and the other is manipulating acoustic features, such as spectrum augmentation. Traditional spectrum augmentation has achieved great success in many tasks. But its time domain mask or frequency domain mask often mask important speech information. This will lead to the introduction of noise in model training and affect the results. GridMask is an effective method in CV field, but it has never been tried in speech field. We first utilize the GridMask in ASCs to alleviate this problem. It also is a method of information deletion. This method is implemented by randomly discarding an area on the feature. However, we can control the size of the mask to retain useful information by GridMask. The effect is equivalent to adding a regular term to the network to avoid network overfitting [12] . Compared with changing the network structure, this method only needs to be augmented during data input, which is simple and convenient.

In many scenarios, the adjustment of model structure is generally by manual. This is not only time-consuming and labor-intensive, but also the results are often sub-optimal. A recent study giving new solution to this problem is AutoML. As one of the mainstream directions of deep learning in recent years, AutoML [17] aims to avoid a large number of manual interventions in the application of machine learning. These manual tunings are manifested in various aspects of machine learning such as feature extraction, model selection, and parameter adjustment [18, 19]. The main purpose of AutoML is to allow the model to automatically learn the appropriate parameters and configurations without manual intervention

[20, 21] . Based on these advantages of AutoML, we also apply it to our scheme for ASCs task.

We will introduce the TC-SKNet, GridMask(GM) and AutoML(AM) in Section 2. In Section 3, we describe the experiments and show the results in Section 4. Finally, conclusions are made in Section 5.

## 2. Research Methodology

In this section, we mainly introduce the scheme we adopted. In Fig 1, three modules of our proposed scheme are shown. They are data augmentation (GridMask), TC-SKNet and AutoML. First, we use AutoML to search for parameters such as learning rate, kernel size, channel, etc. on TC-SKNet to obtain the optimal and low-complexity model parameters. After the model is fixed, AutoML is used again to search for the enhanced parameters of GridMask, and finally the optimal classification result is obtained.

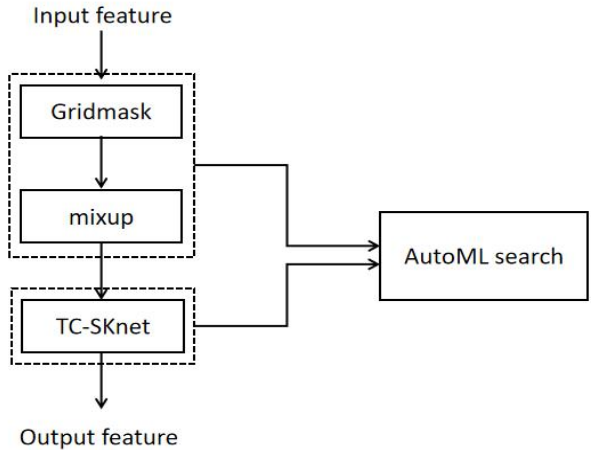

**Fig 1.** The proposed acoustic scene classification system

### 2.1. TC-SKNet

Our new model structure TC-SKNet combines the advantages of temporal-convolution [16] and the Selective Kernel Networks [15]. Temporal-convolution reduces the MACs (Multiply-and-Accumulate Operations) of the model, and Selective Kernel Networks enhances the expressive ability of the model. TC-SKblock is plug-and-play and can be easily embedded into other models and tasks.

The structure of TC-SKNet is shown in Fig 3(b). TC-SKNet consists of two TC-SKblocks, two pooling layers, one 1D convolution layer, Global Average Pooling(GAP) and fully connected layers(FC).

The two TC-SKblocks shown in the Fig 3(a) are similar to SKblock . For any given feature map $X \in R^{T \times C}$, we first conduct two transformations F1 : X→U1 $\in R^{T \times C1}$ and F2 : X→U2 $\in R^{T \times C1}$ with Temporal-Convolution kernel sizes 3 and 5, respectively. After two convolution operations, 1D Batch Normalization and ReLU function in sequence. Second, the results from two branches are fused via an element-wise summation [11]:

$$U = U_1 + U_2 \quad (1)$$

In the next stage, like SEnet[22], we process the information by simply using Global Average Pooling(GAP) [23] to generate channel-wise statistics:

$$S = GAP(U) \quad (2)$$

$S \in R^{C1}$, S undergoes dimensionality reduction through a Fully Connected (FC) layer to obtain Z, and through two FC layers, it is restored to the same dimension as S, $A \in R^{C1}, B \in R^{C1}$. Specifically, a softmax operator is applied on the channel-wise digits:

$$a[i] = \frac{e^{A[i]}}{e^{A[i]} + e^{B[i]}}, b[i] = \frac{e^{B[i]}}{e^{A[i]} + e^{B[i]}} \ i \in C_1 \quad (3)$$

a and b denote the soft attention [14] vectors for U1 and U2. The final feature map F is obtained through the attention weights on various kernels:

$$F[i] = a[i] \times U_1[i] + b[i] \times U_2[i] \ i \in C_1 \quad (4)$$

In the end, we think TC-SKNet can be extended not only to two branches, but also to multiple branches by Eqs. (1) (2) (3) (4).

### 2.2. GridMask

GridMask is a simple, general and efficient strategy, as shown in figure 2. Given an input feature, our algorithm randomly removes some pixels of it. The effect is equivalent to adding a regular term to the network to avoid network overfitting. GridMask It is a method of information deletion. The core requirement of this type of method is to avoid excessive deletion and retention of continuous areas . The current frequency coverage and time domain coverage voice enhancement solutions proposed by Google may cause excessive deletion of one or several regions [12], which may result in the complete deletion of the object and context information, resulting in the remaining information being insufficient for classification and enhancement. The latter data is more like noisy data.

At the same time, too many reserved areas will make some objects inaccessible, and they are characteristic data that may reduce the robustness of the network.

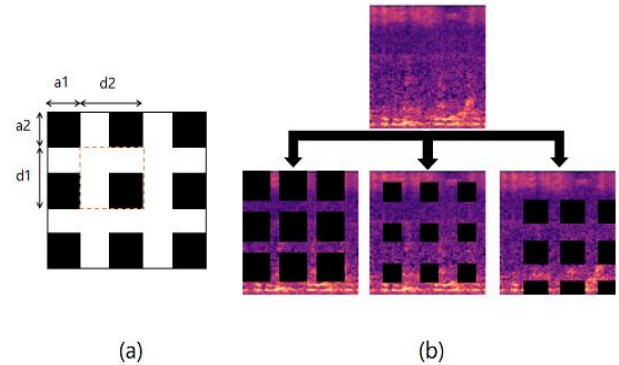

**Fig 2.** GridMask diagrammatic sketch (a) The dotted square shows one unit of the mask. (b) Examples of GridMask with different MR.

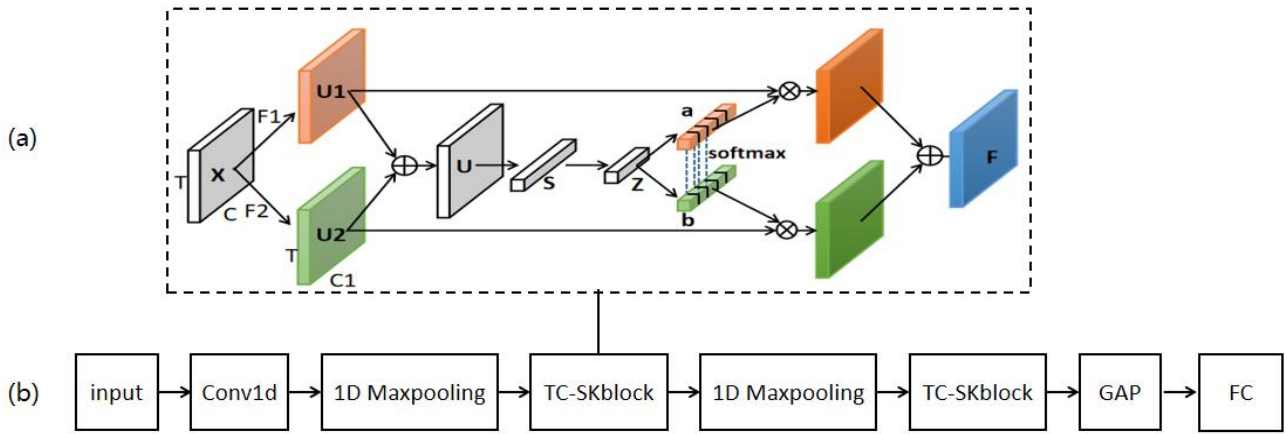

**Fig 3**. (a) TC-SKblock architecture (b) TC-SKNet architecture

GridMask can solve these two problems faced by the information deletion method to a certain extent. To simplify the operation of GridMask, we have defined two hyper-parameters, P (Probability) and MR (Masking ratio). P represents the probability of using augmentation. MR represents the ratio of masking in a region. Based on fig 2 (a), it can be expressed as:

$$MR = (a1 \times a2)/(d1 \times d2) \qquad (5)$$

By adjusting MR, we avoid excessive deletion of noisy data and excessive retained data, which reduces the robustness of the model.

### 2.3 AutoML

The application of machine learning requires a lot of manual intervention, which is manifested in various aspects of machine and evaluation, so that machine learning models can be applied without manual intervention [20]. It can be introduced from two perspectives. From a machine learning perspective, AutoML can be seen as a system with very powerful learning and generalization capabilities on given data and tasks. But it emphasizes that it must be very easy to use.

From the perspective of automation, AutoML can be seen as designing a series of advanced control systems to operate the machine learning model, so that the model can automatically learn the appropriate parameters and configurations without manual intervention. In our work, we adopted the Tree-structured Parzen Estimator Approach (TPE) optimization search method [21] to complete the TC-SKNet parameter search (learning rate, channel, decay) and the GridMask parameter (P, MR) search.

In our experiment, AutoML performs model search on TC-SKNet with a Hyperparameter preset baseline approximation. Accuracy is approximately 0.95% higher than the baseline TC-SKNet and the parameter of model decreased from 34K to 20.9K

### 3. Experiments

In this section, we describe the data and experimental methods.

### 3.1 Data

The development dataset of TAU Urban Acoustic Scenes 2020 Mobile [24] is used to verify the effectiveness of our method. (The evaluation dataset is not published). The recordings in the dataset were collected from 10 acoustic locations. And 3 real devices and 6 simulated devices were used for the dataset.

The total number of recordings is 23,040. The dataset is split into training and test set with a 70% ratio as a cross-validation setup (some recordings are not used for training/test split). The number of training and test set is 13,962 and 2,970, respectively. The recordings from 3 simulated devices are included only in the test set. The duration of each training and test recordings is fixed at 10 sec. The sampling rates are fixed at 44.1 kHz, 24-bit resolution, and mono channel.

### 3.2 Experimental configurations

For each input acoustic signal, a short-time Fourier transform with 1024 FFT points is performed with a hop length of 512 samples. We then extract a 39-dimensional MFCC spectrogram. The number of time bins varied according to the duration of the input acoustic signal, and a feature of 39 × the number of time bins was generated. In model training, we use Adam optimizer with the initial learning rate of 0.001, and a weight decay of 0.0005 and each training 5 epochs is multiplied by the decay 0.98. We also trained the proposed model using a standard cross-entropy loss function with a batch size of 16 and epoch 00..

Table 1: AutoML search space (Model and GridMask).

| Model Hyperparameter search | | |
|---|---|---|
| parameter | type | space |

| parameter | type | space |
|---|---|---|
| learning rate | choice | [0.0001, 0.001, 0.01] |
| batch size | choice | [16, 32, 64, 128, 256] |
| L-size | choice | [25,30,35,40,45,50] |
| C-channels | choice | [30,40,50,60] |
| P-size | choice | [9,11,13,15,17] |
| dropout | uniform | [0.1,0.4] |
| **GridMask Hyperparameter search** | | |
| parameter | type | space |
| P | uniform | [0.5, 1] |
| MR | uniform | [0.1, 0.5] |

AutoML search space shows as Table 1. The first column represents the parameter to be searched, second column represents the search space type (choice: numeric, uniform: continuous interval type), and third column represents the search space. In the Model Hyperparameter search, search for 6 hyperparameters. All of Convolution channels (C-channels) in TC-SKNet, Linear layer sizes in TC-SKblocks (L-sizes). In order to reduce the amount of parameters, the search space of these hyperparameters setting is smaller than the baseline. For the learning rate, batch size and dropout, we set the search space according to the training experience of the previous model. In the searching of hyperparameters P and MR by GridMask, the type of search space is uniform.

Compared with the state-of-the-art CNNs, we designed baseline model with hyperparameters, as presented in Table 2. The size of 27K is smaller than other state-of-the-art CNNs (34.4K-35K)[10,11].

## 4. Results

In this section, the detailed experimental results are shown. TC-SKNet is compared with MobNet and others. Finally, we show TC-SKNet, GridMask, and AutoML on the experimental results in ASCs.

Table 2: Model and GridMask hyperparameters for the baseline and AutoML

| | TC-SKNet(baseline) | TC-SKNet(AM) |
|---|---|---|
| Params (K) | 28K | 20.9K |
| batch size | 16 | 12 |
| learning rate | 0.001 | 0.003 |
| C-channel | 60 | 40 |
| L-size | 50 | 45 |
| P-size | 11 | 15 |
| dropout | 0.2 | 0.145 |
| | GridMask(baseline) | GridMask(AM) |
| P | 0.6 | 0.52 |
| MR | 0.3 | 0.31 |

For the first experiment, we compare GridMask with traditional spectrum augmentations. The results from Table 3 show that compared with traditional spectrum augmentations [25], GridMask can bring an improvement of 0.95%. In other words, the performance gain brought by GridMask is stronger than spectrum augmentation in ASCs.

Table 3: Ablation study of the proposed model using GridMask and spectrum augmentation

| Model | Params(K) | ACC(%) |
|---|---|---|
| TC-SKNet(baseline) | | 53.98 |
| w/ GridMask | 27K | 55.17 |
| w/ spectrum augmentations | | 54.22 |

GridMask and mixup are applied to the proposed baseline model in second experiment, as listed in Table 4. Based on the results of the experiment, it was confirmed that the use of GridMask with mixup showed an absolute performance difference of 4.23%.

We mainly show the performance improvement of AutoML compared with baseline in the third experiment. AutoML search 8 hyperparameters results in Table 2. As can be seen from the Table 3, TC-SKNet with AutoML showed 54.93% accuracy with only 20.9K model parameters. GridMask with AutoML improves our performance to 57.92% is better than GridMask (baseline). Similar to the second experiment, we also adjusted the order of GridMask and mixup. The proposed TC-SKNet achieved a peak accuracy of 59.87% with a model parameter of 20.9K.

Similar to the first experiment, we also adjusted the use order of GridMask and mixup in second. The second experiment and third experiment have the same conclusion, the effect of using GridMask first and then mixup is better than that of using GridMask and mixup. The reason for this phenomenon may be that after mixing data with mixup, GridMask loses the proportion information of two mixed voice at the same time, resulting in performance degradation.

Table 4: Model hyperparameters for the baseline and AutoML

| Model | Params(K) | ACC(%) |
|---|---|---|
| TC-SKNet(baseline) | | 53.98 |
| w/ GridMask(baseline) | | 55.17 |
| w/ mixup($\alpha = 0.2$) | 27K | 56.27 |
| w/ GridMask-mixup | | **58.21** |
| w/ mixup-GridMask | | 52.43 |
| TC-SKNet(AM) | | 54.93 |
| w/ GridMask(baseline) | | 55.67 |
| w/ GridMask(AM) | 20.9K | 57.92 |
| w/ mixup($\alpha = 0.2$) | | 56.44 |
| w/ GridMask(AM)-mixup | | **59.87** |
| w/ mixup-GridMask(AM) | | 53.61 |

The final experiment is to compare with state-of-the-art CNNs in ASC. In the Table 5. Our methods' performance is comparable to those of state-of-the-art CNNs with accuracies ranging from 51.25% to 61.83% and model parameters ranging from 34.4 K to 35 K.

Table 5: Performance comparison between the proposed model and state-of-the-art CNNs.

| Model | Params (K) | ACC(%) |
|---|---|---|
| DCASE 2021 task 1a baseline | 46.2 | 46.4 |
| Residual CNN [10] #num. of stacks & filters = 2& 10 | 34.4 | 51.25 |
| MobNet [11] # num. of filters = {10, 14, 18} | 35 | **61.83** |
| Small-FCNN [11] # num. of filters = {14, 26, 38} | 34.5 | 56.64 |
| TC-SKNet(baseline) | 20.9 | **58.21** |
| TC-SKNet(AM) | 20.9 | **59.87** |

## 5. Conclusions

In this work, we explored new model architectures, augmentation, and AutoML for ASCs. Experiments have proved that our proposed TC-SKNet is a lightweight architecture. Compared with other model architectures, it can reduce the number of parameters while keeping performance, and this architecture can be easily embedded into other models and tasks. By adjusting the parameter MR, you can retain useful information and delete useless information, thereby enhancing the generalization of the model. AutoML can automatically help the model to complete parameter adjustments, which greatly saves our labor costs and can further improves the model performance, but it also faces the risk of overfitting. In general, our solution is simple and easy to implement, suitable for a variety of scenarios, and can be translated to a variety of voice tasks such as speaker verification, wake-up, etc.